\documentstyle[psfig]{elsart}

\begin{document}
\begin{frontmatter}
\title{Super-paramagnetic clustering of yeast gene expression profiles}

\author{G. Getz$^a$, E. Levine$^a$, E. Domany$^a$
and M. Q. Zhang$^b$}

\address{$^a$ Department of Physics of Complex Systems,
Weizmann Institute of Science,
Rehovot 76100, Israel \\
$^b$ Cold Spring Harbor Laboratory,
P. O. Box 100,
Cold Spring Harbor, New York 11724, USA}

\begin{abstract}
High-density DNA arrays, used to monitor gene expression at a genomic scale,
have produced  vast amounts of information which require the development
of efficient computational methods to analyze them. The important first step
is
to extract the fundamental patterns of gene expression inherent in the data.
This paper describes the application of a novel clustering algorithm, 
Super-Paramagnetic Clustering (SPC) to analysis of gene expression
profiles
that were generated recently during a study of the yeast cell
cycle. SPC was used to organize genes into biologically relevant clusters
that
are suggestive for their co-regulation. Some of the advantages of SPC
are its robustness against noise and initialization, a clear signature
of cluster formation and splitting, and an unsupervised self-organized
determination of the number of clusters at each resolution. Our analysis
revealed interesting correlated behavior of several groups of genes 
which has not
been previously identified.
\end{abstract}
\end{frontmatter}

\section{Introduction}

DNA microarray technologies have made it straightforward to monitor
simultaneously the expression levels of thousands of genes during
various cellular processes \cite{Lockhart96}\cite{DeRisi97}. The new
challenge is  to make sense of such massive expression data
\cite{Zhang99}.
In most such experiments, investigators  compare the relative
change of gene expression levels between two samples (one is called the
target,
such as a disease sample; the other is called the control, such as a
normal sample). 
In a typical experiment 
simultaneous expression levels of thousands of genes are viewed
over a few tens of time-points (or different tissues \cite{Alon99}).
Hence one needs to analyse arrays that contain $10^5 - 10^6$ measurements.

The aims of such analysis are typically to
(a) group genes with correlated expression profiles;
(b) Focus on those groups which seem to participate in some biological process;
(c) Provide a biological interpretation of the clusters. 
Interpretations could be 
co-regulation of the mean cluster expression with a known process,
a promoter common to most of the genes in the cluster, etc.
(d) In experiments that compare data from different tissues 
(such as tumor and normal~\cite{Alon99}) one also tries to
differentiate them  on the basis of their genetic expression profiles.

The sizes of the datasets and their complexity call for 
multi-variant
clustering techniques which are essential for extracting correlated patterns
in the swarm of data points in multidimensional space (for example,
each relative
gene
expression profile with k time-points may be regarded as a k-dimensional
vector).
\section{SPC }

Currently, two clustering appoaches are very popular among biologists. One
is average linkage, a
hierarchical clustering method \cite{Hartigan75}, 
with the Pearson correlation used as a similarity measure
\cite{Eisen98}. The other is self-organizing maps (SOMs) \cite{Kohonen97},
whose most popular implementation for array data analysis is GENECLUSTER
\cite{Tamayo99}.

We present here clustering performed by SPC, 
 a hierarchical clustering method recently introduced by 
Blatt et al \cite{Blatt96a}. 
It is based on an analogy to the physics of inhomogeneous ferromagnets.
Full details of the algorithm~\cite{ncomp} and the
underlying philosophy~\cite{statphys} are given elsewhere
; here only a brief
description is provided.

The input required for
SPC is a distance matrix between the $N$ data points that are to be clustered.
From such a distance matrix one constructs 
a graph,
whose vertices are the data points and edges identify neighboring points. 
Two points $i$ and $j$ 
are called neighbors (and connected by an edge) if they satisfy the
{\it K-mutual-neighbor} criterion, i.e.  iff $j$ is one of the $K$ nearest
points to $i$ and vice versus. With each edge we associate a weight $J_{ij}>0$,
which decreases as the distance between points $i$ and $j$ increases.

Assignement of the datapoints to clusters is equivalent to partitioning this
weighted graph.
Cluster indices play the role of the states of
Potts spins assigned to each vertex (i.e. to each original data point).
Two neighboring spins are interacting ferromagnetically
with  strength $J_{ij}$.
This Potts ferromagnet is simulated at a sequence of temperatures $T$. The
susceptibility and the correlation function for neighboring pairs of spins
are measured. The pair correlation function serves to identify clusters: 
high correlation means that the two data points belong to the same cluster.

The temperature $T$ controls the resolution at which  
clustering is performed; 
the algorithm finds typical clusters at all resolutions.
At very low temperatures all points belong to a single cluster
and as $T$ is increased, clusters break into smaller ones until 
at high enough temperatures each point forms its own cluster.
The clusters found at all temperatures form a dendrogram.
Blatt et al showed that the SPC algorithm is robust since the clusters
are formed due to collective behavior of the system. 
The major splits can be 
easily identified by a peak in the susceptibility. 
For more details see \cite{Blatt96a,ncomp,statphys}.

\section{Yeast Cell Cycle and Microarray Data}
We applied SPC on a recently published data set~\cite{web}
to determine whether it could
automatically expose known clusters without using prior knowledge. 
Eisen et al \cite{Eisen98} clustered the genes on the basis of data
{\it combined} from 7 different experiments. 
We suspected that mixing the results of different experiments
may introduce noise into the data associated with a single one. Therefore
we chose to use only a single time course, that of 
gene expression as measured in a single process (cell division cycles
following alpha-factor
block-and-release~\cite{Spellman98}). Furthermore, we 
focused on genes that have characterized functions (2467 genes) for easier
interpretation.

Genetic controls and regulation play a central role in determination of cell
fate during development. They are also important for the timing of cell
cycle
events such as DNA replication, mitosis and cytokinesis. Yeast is a single
cellular organism, which has become a favorite
model
in molecular biology
due to the easiness of genetic and biochemical
manipulation
and the availability of the complete genome. 
Like all living cells, the yeast cell cycle consists of four phases:
G1$\rightarrow$S$\rightarrow$G2$\rightarrow$M$\rightarrow$G1..., where
S is the  phase
of DNA synthesis (replicating the genome); M stands for mitosis (division
into two daughter cells), and the two gap phases are called G1 (preceding the 
S phase) and
G2 (following the S phase). 
At least four different classes of cell cycle regulated
genes exist in yeast \cite{Koch94}: G1 cyclins and DNA synthesis genes are
expressed in late G1; histone genes in S; genes for transcription
factors,
cell cycle regulators and replication initiation proteins in G2; and genes
needed for cell separation 
are expressed as cells enter G1. Early and late G1-specific
transcription is mediated by the Swi5/Ace2 and Swi4/Swi6 classes of factors,
respectively. Changes in the master cyclin/Cdc28 kinases are involved in all
classes of regulation.

In the alpha-factor block-release experiments, MATa cells were first
arrested
in G1 by using alpha pheromone. Then the blocker was removed; from this 
point on the cell division cycle starts and the population 
progresses with significant cell cycle synchrony. RNA was extracted from the
synchronized sample, as well as a control sample (asynchronous cultures of the same cells
growning exponentially at the same temperature in the same medium).

Fluorescently labeled cDNA was synthesized using Cy3 ("green") for the
control
and Cy5 ("red") for the target. Mixtures of equal amounts of the two samples
were taken at every 7min and competitively hybridized to individual
microarrays
containing essentially all yeast genes. The ratio of red to green light
intensity (proportional to the RNA concentrations) was measured by scanning
laser microscopy (See \cite{Spellman98} for experimental details). The
actual
data provided at the Stanford website~\cite{web} is the log ratios.

In the their analysis, Spellman et al. were focusing on identification of
800
cell cycle regulated genes (that may have periodic expression profiles). In our
test of SPC, in addition to oscillatory genes we were
also looking for any groups of genes with highly correlated
expression patterns.

\section{SPC Analysis of Yeast Gene Expression Profiles}

We clustered the expression profiles of the 2467 yeast genes of known function
over data taken at 18 time intervals (of 7 min) 
during two  cell division cycles, 
synchronised by alpha arrest and release. Denote by $E_{ij}$ the relative
expression of gene $i$ at time interval $j$. 
Our data consist of 2467 points
in an 18-dimensional space, normalised in the standard way:
\begin{center}
$G_{ij}=\frac{E_{ij}-<E_i>}{\sigma_i}; \qquad
<E_i>=\frac{1}{18}\sum_{j=1}^{18}E_{ij}; \qquad 
\sigma_i^2=\frac{1}{18}\sum_{j=1}^{18}E_{ij}^2 - <E_i>^2
$
\end{center}
We looked for clusters of genes with correlated expression profiles over the 
two division cycles. The SPC algorithm was used with $q=20$ component Potts
spins, each interacting with those neighbors that satisfy the 
 $K$-mutual neighbor criterion\cite{ncomp} with $K=10$. 
 Euclidean distance between the normalized vectors 
was used as the distance between
two genes. This distance is proportional to the Pearson correlation used
by Eisen {\it et. al.}.  

\begin{figure}[htbp] 
\centerline{
    \psfig{figure=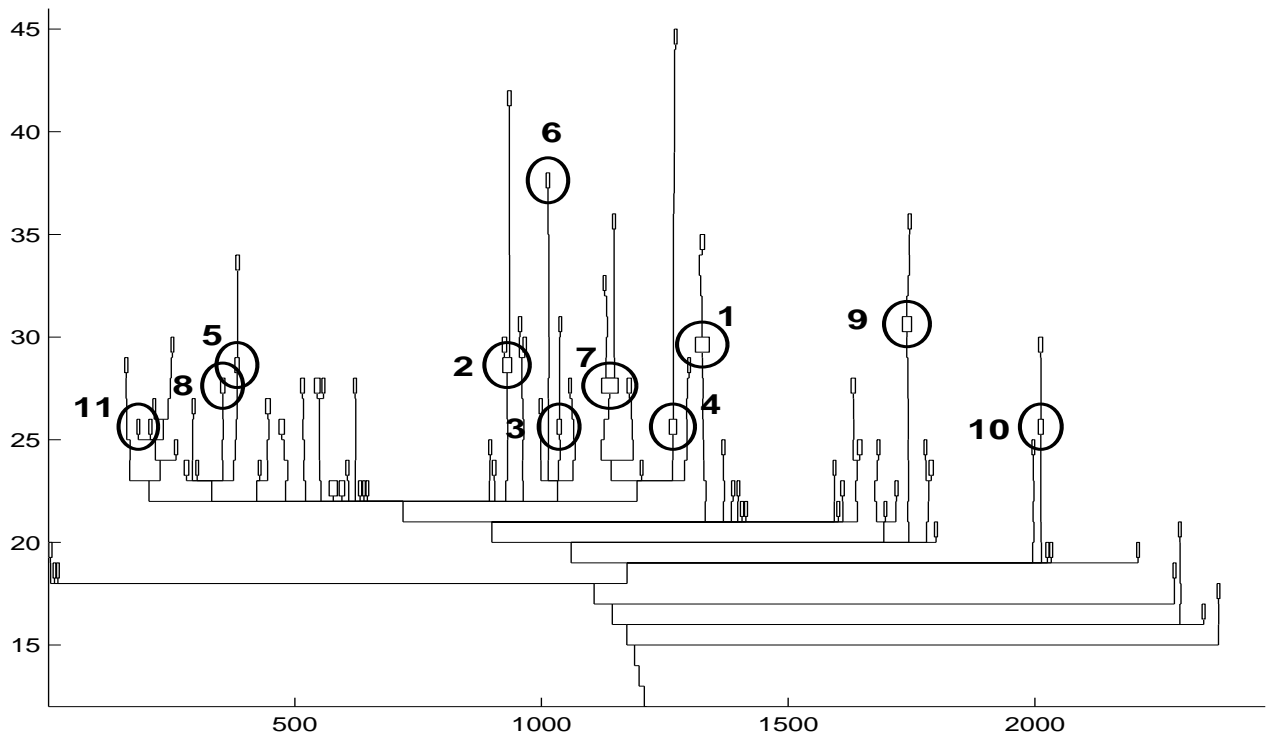,height=10.0cm,width=17.0cm}
}
    \caption{Dendrogram of genes. Clusters 1 - 4 were selected according to our
    criteria, eq. (1 - 2). The other
    circled and numbered clusters are also interesting (see text).} 
    \label{fig:dend}
\end{figure}

At $T=0$ all datapoints form one cluster, which splits as the system
is ``heated". The resulting dendrogram of genes is presented in Fig.
\ref{fig:dend}. Each node represents a cluster; only
clusters of size larger than 6 genes are shown. The last such clusters of
each branch, as well as non-terminal clusters that were 
selected for presentation and analysis
(in a way described below) are shown as boxes. 
The circled boxes represent the clusters that are analysed below. 

The position of every node along the horizontal axis is 
determined for the corresponding cluster according to 
a method introduced by Alon et al~\cite{Alon99}; proximity of two clusters along 
this axis indicates that the corresponding temporal expression profiles are not 
very different.
The vertical axis represents the resolution, controlled by the ``temperature"
$T \geq 0$. The vertical position  of a
node or
box is determined by the value of $T$ at which it splits. 
 A high vertical position indicates that the cluster is 
stable, i.e. contains a fair number of closely-spaced data points (genes with
similar expression profiles). 
 
In order to identify clusters of 
genes whose temporal variation is on the scale
of the cell cycle, we calculated {\it for each cluster} a cycle score $S_1$,
defined as follows. First, for each cluster $C$ (with $N_C$ genes)
we calculate the average normalized expression level at 
all $j=1,...,18$ time intervals and the 
corresponding standard deviations $\sigma^C(j)$:
\[
\bar{G}^{C}(j)=\frac{1}{N_C}\sum_{i \in C} G_{ij} \qquad \qquad
[\sigma^C(j)]^2=\frac{1}{N_C}\sum_{i \in C} (G_{ij})^2  - [\bar{G}^{C}(j)]^2
\]
Next,
we evaluated the Fourier transform of the mean expression profiles 
$\bar{G}^{C}(j)$ for
every gene cluster $C$. To suppress initial transients, 
the Fourier transform is performed 
only over $j=4,...,18$.
Denote the absolute values of the Fourier
coefficients by $A_k$; the ratio between 
low-frequency coefficients and the 
high-frequency ones 
was used as a figure of merit for the time scale of the variation.
We observed that clusters that satisfy the condition
\begin{equation}
S^C_1=\frac{\sum_{k=2}^{4}{A_k}}{\sum_{k=6}^{8}{A_k}} > 2.15
\end{equation}
have the desired time dependence,
and found 29 clusters (consisting of 167 genes) to have such scores.
For many of these clusters, however, the temporal variation was very weak,
i.e.  of the same order
as the standard deviations $\sigma^C(j)$
of the individual gene expressions of the cluster.
We defined another score, $S_2^C$, for which we required
\begin{equation}
S_2^C=\frac{1}{18}\sum_{j=1}^{18}\left[ {\frac{\bar{G}^{C}(j)}{\sigma^C(j)}}
\right]^2
> 5.6
\end{equation}
For clusters $C$ that satisfy this condition the ``signal'' 
significantly exceeds the noise.
We select a cluster if its score exceeds 5.6, 
while its parent's score does not.
Only 4 clusters, containing 86 genes, 
satisfy both conditions (1) and (2); these are 
numbered 1 -- 4 on Fig. \ref{fig:dend}. 
Seven additional relatively stable clusters which did {\em not} 
satisfy our two  criteria, but are of
interest, are also selected
and circled on figure \ref{fig:dend}.

\begin{figure}[htbp] 
\centerline{
    \psfig{figure=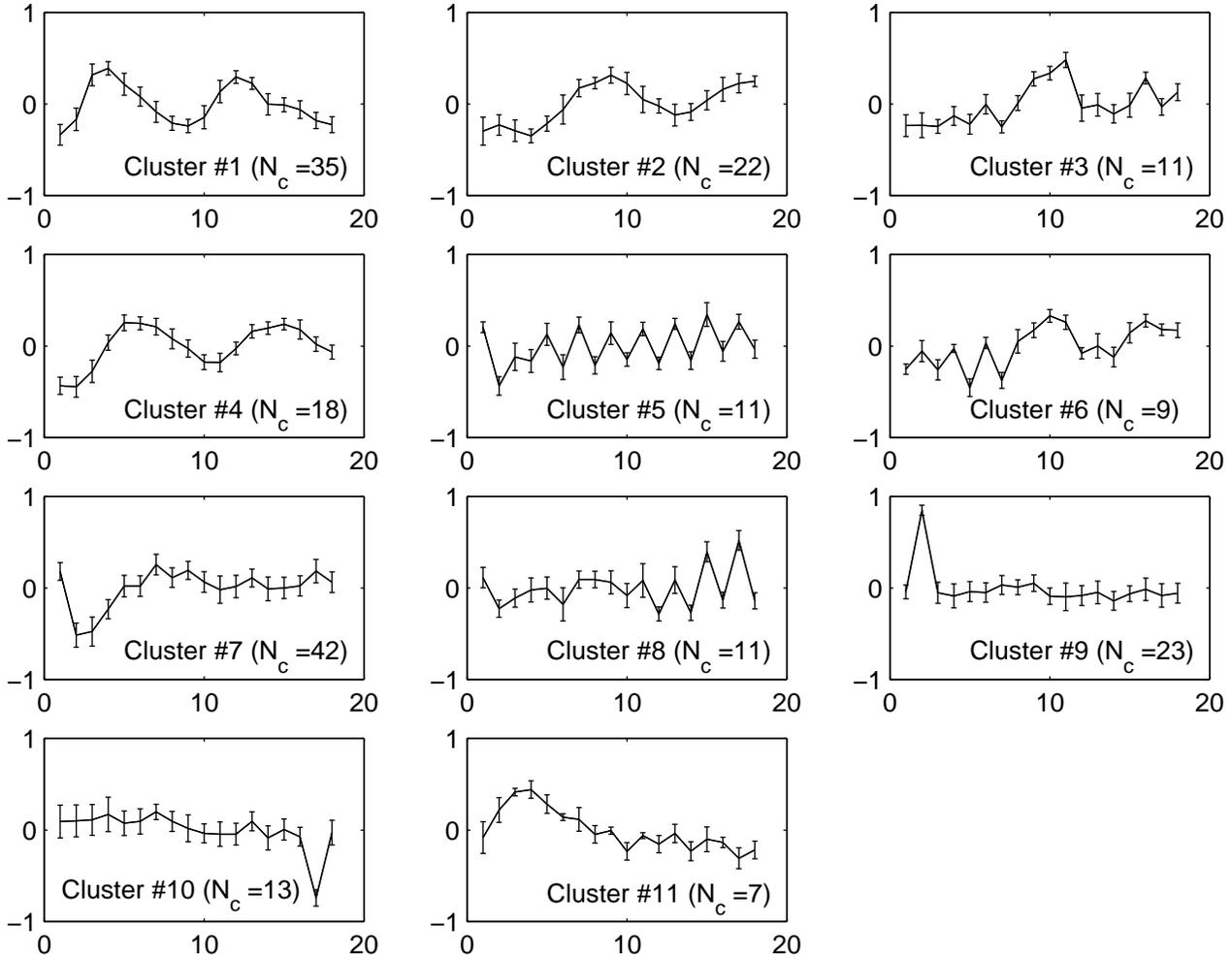,width=17.0cm}
}
    \caption{Mean normalized expression of selected clusters, versus time,
measured at intervals of 7 minutes.    
    Error bars represent the standard
deviations $\sigma^C(j)$. $N_c$ is the number of genes in each cluster.
 The clusters
are numbered as in figure 1.}
    \label{fig:seq}
\end{figure}

The corresponding time sequences are shown
in Fig \ref{fig:seq}: 
$ \bar{G}^{C}(j)$ is plotted for each cluster versus time $j$,
with the error bars representing the standard deviations $\sigma^C(j)$.
Clusters 1,2 and 4 clearly corresponds to the cell cycle. 

\section{Details and Interpretation of gene clustering.}

The full lists of genes that constitute the 11 selected clusters 
are given in our website~\cite{ourweb}. We present here a short 
analysis of our clusters. We use standard notation for bases:  R stands for 
A or G, W for A or T, K for G or T, N for any base.

{\bf Cluster \# 1:}
These are mostly Late G1 phase specific genes. They contain the major cell cycle
regulators: Cln1,2, Clb5,6 and Swi4 as well as DNA replication and repair genes.
One can easily detect MCB (ACGCGT) or SCB (CRCGAAA) sites in their promoters to
which MBF (Swi6p+Mbp1p) and SBF (Swi6p+Swi4p) bind respectively \cite{Koch94}.

{\bf Cluster \# 4:}
This cluster contains mostly S phase genes and is dominated by the histones.
Histones are required for wrapping up nascent DNA into nucleosomes, their
promoters are regulated by CCA (GCGAARYTNGRGAACR), NEG (CATTGNGCG) as well
as SCB (CGCGAAA) \cite{Zhang99}.

{\bf Cluster \# 2:}
These are mostly G2/M phase genes. They contain the major cell cycle regulators:
Clb1,2 and Swi5/Ace2. It is known that all genes co-regulated with Clb1,2 are
mainly controlled by either  Mcm1 at P-box (TTWCCYAAWNNGGWAA) or by Mcm1+SFF
through the composite site: (P-box)N2-4RTAAAYAA \cite{Spellman98}\cite{Zhang99}.

{\bf Clusters \# 5, \# 6 and \# 8:}
These are mostly ribosomal protein (RP) genes. The
genome of Saccharomyces cerevisiae contains 137 
genes coding
for ribosomal proteins \cite{Mager97}. Since 59 genes are duplicated, the
ribosomal gene family encodes 78 different proteins, 32 of the small and
46 of the large ribosomal subunit. They are co-regulated because
they are sub-components of ribosome machinery for protein translation. All genes
in cluster \#6 reside on chromosomes 2, 4 and 5, except rpl11b which resides
on chromosome 9. All genes in clusters \#5 and \#8 (which are very close 
in the
dendrogram of Figure 1) reside on chromosomes 8-16, except
rps17b which resides on chromosome 4. It is likely that the expression of
these ribosomal genes is correlated to their chromosomal locations.  It is
interesting that the expression profiles appear to have 
pronounced oscillations
(throughout in \#5, at early times in \#6 and late times in \#8). 
Like most of the
RP genes, the ribosomal genes in the 3 clusters also contain multiple global
Regulator Rap1p binding sites in their promoters within a preferred window of
15-26 bp \cite{Lascaris99}. The transcription of most RP genes is activated by 
two Rap1p binding sites, 250 to 400 bp upstream from the initiation of 
transcription. Since Rap1p can be both an activator and a silencer, it is
not known whether Rap1p is responsible for the oscillation. This oscillation
could be a result of interplay between cell cycle and Rap1p activity which
determines the mean half life of the RP mRNAs (5-7min, \cite{Li99}). As fresh
medium was added at 91min during the alpha-factor experiments, the genes
in \#6 and in \#8 may have different responses to the nutrient change.

{\bf Cluster \#7:}
This cluster has 42 genes that are largely not cell cycle regulated. These
genes have diverse functions in general metabolism. When searching promoter
regions for regulatory elements using gibbsDNA \cite{Zhang99a}, a highly
conserved motif GCGATGAGNT is shared by 90 \% of genes. 
This element seems to be
novel, it has some similarity to Gcn4p site TGACTC and Yap1p site GCTGACTAATT
\cite{Hinnebusch92}. When searching the yeast promoter database - SCPD 
\cite{Zhu99}, we found that the BUF site in the HO gene promoter and 
the UASPHR site in the
Rad50 promoter appear to contain the core motif GATGAG. Although we do
not know if this element is functional or what might be the trans-factor,
it is still very likely that it may contribute the co-regulation of this
cluster of genes.

{\bf Cluster \#10:}
This cluster is characterized by a pronounced 
dip towards the end of the profile.
They are not cell cycle regulated by and large, except Clb4 (a S/G2 cyclin
) and Rad54 (a G1 DNA repair gene). By searching promoter elements, we found
a conserved motif RNNGCWGCNNC that is shared by a subset of the genes
(Clb4, YNL252C, Rad54, Rpb10, Atp11 and Pex13). It partially matches a
PHO4 binding motif (TCGGGCCACGTGCAGCGAT) in the promoter of Pho8. However, 
the
PHO4 consensus, CACGTK, does not appear in 
the conserved motif of our cluster. 
Therefore we suspect that it is a novel motif which should
be tested by experiments.

\section{Summary}

We used the SPC algorithm to cluster gene expression data for the yeast genome.
We were able to identify groups of genes with highly correlated temporal
variation. Three of the groups found clearly correspond to well known phases of
the cell cycle; some of our observations of other clusters reveal features that
have not been identified previously and may serve as the basis of future
experimental investigations.

\newpage
{\bf Acknowledgements}

Research of E. Domany was partially supported by the Germany-Israel Science
Foundation (GIF) and the Minerva foundation. Research of M. Q. Zhang was
partially supported by NIH/NHGRI under the grant number HG01696.


\begin{thebibliography}{99}
\bibitem{Lockhart96}Lockhart DJ, Dong H, Byrne MC, et al. (1996)
{\em Nat. Biotech.} {\bf 14}, 1675-1680.
\bibitem{DeRisi97} De Risi J, Iyer V and Brown PO (1997) {\em Science}
{\bf 278}, 680-686.
\bibitem{Zhang99} Zhang MQ (1999) {\em Genome Res.} {\bf 9}, 681-688.
\bibitem{Alon99}
Alon, U., Barkai, N., Notterman, D.A., Gish, K.,  Ybarra, S., Mack, D., AND Levine, A. J.
(1999) Proc. Natl. Acad. Sci. 
{\bf 96}, 6745--6750.
\bibitem{Hartigan75} Hartigan J (1975) {\em Clustering Algorithms} (Wiley,
New York).
\bibitem{Eisen98} Eisen M, Spellman PT, Brown PO and Botstein D (1998) {\em
Proc. Natl. Acad. Sci. USA} {\bf 95}, 14863-14868.
\bibitem{Kohonen97} Kohonen T (1997) Self-Organizing Maps (Springer,
Berlin).
\bibitem{Tamayo99} Tamayo P, Slonim D, Mesirov J et al. (1999) {\em Proc.
Natl. Acad. Sci. USA} {\bf 96}, 2907-2912.
\bibitem{Blatt96a} Blatt,~M., Wiseman,~S., and Domany,~E. 1996a.
  ``Super--paramagnetic clustering of data'', {\em Physical Review
    Letters} {\bf 76}, 3251--3255.
\bibitem{ncomp}M. Blatt, S. Wiseman and E. Domany, {\it Neural Computation}
{\bf 9} 1805 (1997).
\bibitem{statphys}E. Domany, Physica {\bf A 263}, 158 (1999)    
\bibitem{Spellman98} Spellman PT, Sherlock G, Zhang MQ et al. (1998) {\em
Mol. Biol. Cell.} {\bf 9}, 3273-3297.
\bibitem{Koch94} Koch C and Nasmyth K (1994) {\em Curr. Biol.} {\em 6},
451-459.
\bibitem{web} The data can be obtained from 
http://cellcycle-www.standford.edu 
\bibitem{ourweb} http://www.weizmann.ac.il/physics/complex/clustering/
\bibitem{Mager97} Mager WH, Planta RJ, Ballesta JG et al. (1997) {\em Nucl. Acid. Res.} {\bf 25}, 4872-4875.
\bibitem{Lascaris99} Lascaris RF, Mager WH and Planta RJ (1999). {\em Bioinformatics} {\bf 15}, 267-277.
\bibitem{Li99} Li B, Nierras CR and Warner JR (1999). {\em Mol Cell Biol} {\bf 19}, 5393-5404.
\bibitem{Hinnebusch92} Hinnebusch AG (1992). In {\em The Molecular and Cellular
Biology of the Yeast Sacchromyces: Gene Expression}, Vol.2, pp319, Cold Spring 
Harbor Press, New York.
\bibitem{Zhang99a} Zhang MQ (1999a). {\em Comput. Chem.} {\bf 23}, 233-250.
\bibitem{Zhu99} Zhu J and Zhang MQ (1999). {\em Bioinformatics} {\bf 15}, 
607-611.

\end{thebibliography}
\end{document}